\documentclass[osajnl,twocolumn,showpacs,superscriptaddress,10pt]{revtex4-1} %% use 11pt for Applied Optics
\usepackage{amsmath,amssymb,graphicx}

\newcommand{\bra}[1]{\mbox{$\left\langle #1 \right|$}}
\newcommand{\ket}[1]{\mbox{$\left| #1 \right\rangle$}}

\begin{document}

\title{Long distance measurement-device-independent quantum key distribution with coherent-state superpositions}

%\author{}
\author{Hua-Lei Yin}\email{Corresponding author: hlyin@mail.ustc.edu.cn}
\author{Wen-Fei Cao}
\author{Yao Fu}
\author{Yan-Lin Tang}
\author{Yang Liu}
\author{Teng-Yun Chen}\email{Corresponding author: tychen@ustc.edu.cn}
\author{Zeng-Bing Chen}\email{Corresponding author: zbchen@ustc.edu.cn}
\affiliation{Hefei National Laboratory for Physical Sciences at Microscale and Department of Modern Physics,\\University of Science and Technology of China, Hefei, Anhui 230026, China}
\affiliation{CAS Center for Excellence and Synergetic Innovation Center of Quantum Information and Quantum Physics, \\
University of Science and Technology of China, Hefei, Anhui 230026, China \\}

\begin{abstract}
Measurement-device-independent quantum key distribution (MDI-QKD) with decoy-state method is believed to be securely applied to defeat various hacking attacks in practical quantum key distribution systems. Recently, the coherent-state superpositions (CSS) have emerged as an alternative to single-photon qubits for quantum information processing and metrology. Here, in this Letter, CSS are exploited as the source in MDI-QKD.  We present an analytical method which gives two tight formulas to estimate the lower bound of yield and the upper bound of bit error rate. We exploit the standard statistical analysis and Chernoff bound to perform the parameter estimation. Chernoff bound can provide good bounds in the long distance MDI-QKD. Our results show that with CSS, both the security transmission distance and secure key rate are significantly improved compared with those of the weak coherent states in the finite-data case.

\end{abstract}

% REPLACE WITH CORRECT OCIS CODES FOR YOUR ARTICLE
                          % NOTE: \ocis{} IS ALIASED TO \pacs{} BUT MUST
                          % FORMAT THE TERMS CORRECTLY FOR EACH JOURNAL
\maketitle

Quantum key distribution (QKD) \cite{BB_84,ekert1991quantum} provides the guarantee for two remote parties to share a secret key with information-theoretic security based on the principles of quantum mechanics \cite{Shor:Simple:2000}, despite of the existence of eavesdroppers.
However, there are many imperfections in the realistic QKD systems related to the practical devices failing to satisfy the assumptions in the security proof. By means of these imperfections, various quantum hacking strategies have been proposed to attack the existing QKD systems \cite{zhao:Quantum:2008,Lydersen:BrightAttack:2010,weier2011quantum}. Recently, a novel idea of measurement-device-independent QKD (MDI-QKD) has been proposed \cite{Lo:MIQKD:2012}, in which the protocol is naturally immune to all side-channel attacks on the detectors.
Since the extensively adopted WCS contains multi-photon components, the decoy-state method \cite{Lo:Decoy:2005,wang2005beating} is used in the MDI-QKD to avoid photon-number-splitting (PNS) attacks \cite{Brassard:PNS:2000}. Recently, several experimental demonstrations of MDI-QKD with weak coherent states (WCS) have been performed \cite{Liu:MDIQKD:2013,Tittel:MDIQKD;2013,Tang:2013:exp,da2013proof,Tang:2014:MDIQKD}. Up to now, the maximal experimental transmission distance is 200 km through spooled standard telecom fiber using the decoy-state MDI-QKD \cite{Tang:2014:MDIQKD}.
Nevertheless, the implementation of long distance is still a big challenge for QKD.
Actually, since the WCS sources contain a large portion of vacuum state and multi-photon components, the transmission distance and the secure key rate are limited.

The coherent-state superpositions (CSS) have recently emerged as an alternative to single-photon qubits for quantum information processing and metrology, such as fault-tolerant linear optical quantum computation \cite{Lund:Fault:2008}, quantum teleportation \cite{andersen2013high,PhysRevA.64.052308,PhysRevA.64.022313}, quantum repeaters \cite{sangouard2010quantum}, hybrid long distance entanglement distribution \cite{Brask:Hybrid:2010} and quantum precision measurements \cite{PhysRevA.66.023819}.
Furthermore, approximate CSS of small amplitudes have been generated by photon subtraction from a squeezed vacuum state \cite{neergaard2006generation}, and approximate CSS of large amplitudes have been generated from Fock states using a single homodyne detection \cite{ourjoumtsev2007generation}.

The CSS are often called Schr\"odinger cat states which are defined as quantum superpositions of classical distinguishable states. %Due to its many attractive properties,
It can be written as
\begin{equation} \label{coherent state superposition shown}
\begin{aligned}
\ket{\psi} &= \frac{1}{\sqrt{2(1-e^{-2|\alpha|^2})}}\big(\ket{\alpha}-\ket{-\alpha}\big),
\end{aligned}
\end{equation}
where $\ket{\alpha}$ and $\ket{-\alpha}$ correspond to the coherent states with the same amplitude and opposite phase.
When the phases of CSS pulses sent by Alice and Bob are randomized, the states become a mixture of Fock states.
In practical system, it will have even-photon components for non-ideal CSS \cite{neergaard2006generation}. In general, the density matrix of phase randomized non-ideal CSS can be written as
\begin{equation} \label{CSS}
\begin{aligned}
\sigma=\sum_{i=0}^{\infty}P(i)\ket{i}\bra{i}.
\end{aligned}
\end{equation}
However, we can not know exactly $P(i)$ because the states are non-ideal. Here, we assume the density matrix of phase randomized non-ideal CSS in this Letter can be written as
\begin{equation} \label{CSS}
\begin{aligned}
\rho=&\frac{a}{\sinh{\mu}}\sum_{i=0}^{\infty}\frac{\mu^{2i+1}}{(2i+1)!}{\ket{2i+1}\bra{2i+1}}\\
&+\frac{1-a}{\cosh{\mu}}\sum_{i=0}^{\infty}\frac{\mu^{2i}}{(2i)!}{\ket{2i}\bra{2i}}\\
=&\sum_{i=0}^{\infty}P_{\mu}(i)\ket{i}\bra{i},
\end{aligned}
\end{equation}
$a=1$ represents the CSS case; otherwise it is the non-ideal CSS case. The fidelity of the non-ideal CSS is $\sqrt{a}$, $\mu=|\alpha|^2$ is the intensity of the CSS pulses, and $P_{\mu}(i)$ represents the probability of emerged $n$-photon component in the mixed state.

For simplicity, we consider the  polarization-encoding scheme of decoy state MDI-QKD protocol \cite{Lo:MIQKD:2012}. Alice and Bob independently and randomly send the phase randomized CSS pulses with the intensity $\mu$ and $\nu$ to Charlie under basis $\omega_{A}$ and $\omega_{B}$, respectively, where $\omega_{A},~\omega_{B} \in\{Z,X\}$ ($\mu^{\omega_{A}}\uplus\nu^{\omega_{B}}$ channel).
Charlie performs a Bell-state measurement (BSM), then he employs a public channel to announce the measurement result. Afterwards, Alice and Bob perform basis sift, error correction and privacy amplification to extract a secure key.
The total gain $Q_{\mu\nu}^\omega$ and quantum bit error rate $E_{\mu\nu}^\omega$ (given that Alice and Bob use CSS pulses with $\mu$ and $\nu$ intensity, respectively, and they both choose the same $\omega$ basis) can be written as
\begin{equation} \label{gain and error}
\begin{aligned}
Q_{\mu\nu}^\omega &= \sum_{i,j=0}^{\infty}P_{\mu}(i)P_{\nu}(j)Y_{ij}^{\omega},\\
E_{\mu\nu}^{\omega}Q_{\mu\nu}^\omega &= \sum_{i,j=0}^{\infty}P_{\mu}(i)P_{\nu}(j)e_{ij}^{\omega}Y_{ij}^\omega,
\end{aligned}
\end{equation}
where $\omega \in \{Z,X\}$, $Y_{ij}^{\omega}$ and $e_{ij}^{\omega}$ are the yield and bit error rate given that Alice and Bob send out $i$-photon state and $j$-photon state, respectively. The secure key rate is given by \cite{Lo:MIQKD:2012,ma2012statistical}
\begin{equation} \label{finally key rate}
\begin{aligned}
R &= Q_{11}^Z[1-H(e_{11}^X)]-Q_{\mu\nu}^ZfH(E_{\mu\nu}^Z),
\end{aligned}
\end{equation}
where $Q_{\mu\nu}^ZfH(E_{\mu\nu}^Z)$ is the cost of bit error, $Q_{11}^ZH(e_{11}^X)$ is the cost of phase error, $Q_{11}^Z=P_{\mu}(1)P_{\nu}(1)Y_{11}^Z$ is the yield of single-photon state.
$f=1.16$ is the error correction efficiency, $H(x)=-x\log _{2}(x)-(1-x)\log _{2}(1-x)$ is the binary Shannon entropy function.
\begin{table*}
\centering
\caption{Alice and Bob post-select the successful measurement result when they use the same basis. Here, $\ket{HH}$ represents a state that Alice and Bob both send out horizontal polarization pulse. $\surd$ ($\times$) represents a correct (false) BSM result.} \label{Tab:Suc:BEM}
\begin{tabular}{c||c|c|c|c|c|c|c|c}
\hline
\hline
$ $  & $\ket{HH}$ &   $\ket{VV}$ &   $\ket{HV}$ & $\ket{VH}$ & $\ket{++}$ & $\ket{--}$ & $\ket{+-}$ & $\ket{-+}$  \\
\hline
$\ket{\psi^+}$ & bit flip $\times$ & bit flip $\times$& bit flip $\surd$& bit flip  $\surd$ & no bit flip $\surd$& no bit flip $\surd$& no bit flip $\times$& no bit flip $\times$\\
\hline
$\ket{\psi^-}$ & bit flip $\times$& bit flip $\times$& bit flip $\surd$& bit flip $\surd$& bit flip $\times$& bit flip $\times$& bit flip $\surd$& bit flip $\surd$\\
%\hline
%$\ket{\psi^+}$ & $\times$ & $\times$ & $\surd$ & $\surd$ & $\surd$ & $\surd$ & $\times$ & $\times$ \\
%\hline
%$\ket{\psi^-}$ & $\times$ & $\times$ & $\surd$ & $\surd$ & $\times$ & $\times$& $\surd$ & $\surd$  \\
\hline
\hline
\end{tabular}
\end{table*}
$Q_{\mu\nu}^\omega$ and $E_{\mu\nu}^\omega$ can be directly measured experimentally, while $Q_{11}^\omega$ and $e_{11}^\omega$ need to be estimated with the decoy-state method \cite{Lo:Decoy:2005,wang2005beating}.

First of all, we consider the joint quantum state given that Alice and Bob send out $i$-photon and $j$-photon state with horizontal polarization, respectively, which can be given by
\begin{equation} \label{iniHjH}
\begin{aligned}
\ket{\Psi}_{in}^{HH}=\ket{i}_H\ket{j}_H=\frac{(a_{1H}^\dag)^i}{\sqrt{i!}}\frac{(a_{2H}^\dag)^j}{\sqrt{j!}}\ket{0},
\end{aligned}
\end{equation}
where $a_{1H}^\dag$ and $a_{2H}^\dag$ are photon creation operators with the horizontal polarization mode. After the state goes through the beam splitter and the polarization beam splitter, the output quantum state will be given by
\begin{equation} \label{outiHjH}
\begin{aligned}
\ket{\Psi}_{out}^{HH} = &\sum_{p=0}^{i+j}\sum_{k=0}^{j}\frac{(-1)^{j-k}C_{i}^{p-k}C_{j}^{k}}{\sqrt{2^{i+j}i!j!}}\sqrt{p!}\sqrt{(i+j-p)!}\\
&\times\ket{p}_{1H}\ket{0}_{1V}\ket{i+j-p}_{2H}\ket{0}_{2V},
\end{aligned}
\end{equation}
where $\ket{\Psi}_{out}^{HH}$ can be described as superpositions of orthogonal states $\ket{p}_{1H}\ket{0}_{1V}\ket{i+j-p}_{2H}\ket{0}_{2V}$. Other polarization cases can be considered in the similar manner.

From Table \ref{Tab:Suc:BEM}, it is clear that $Q_{\mu\nu}^{Z}$ contains the correct BSM result $Q_{C}^{Z}$ (a successful BSM given that the  polarization of the pulses sent out by Alice and Bob are different in $Z$ basis) and the false BSM result $Q_{E}^{Z}$, which can be given by
\begin{equation} \label{QCZ}
\begin{aligned}
Q_{\mu\nu}^{Z} &= Q_{C}^{Z}+Q_{E}^{Z},~~Q_{C}^{Z}= 4Q_{HV}^{\psi^+},~~Q_{E}^{Z} = 4Q_{HH}^{\psi^+},
\end{aligned}
\end{equation}
where $Q_{HV}^{\psi^+}$ is the successful Bell state $\ket{\psi^+}$ measurement results given that Alice and Bob send out horizontal and vertical polarization pulse, respectively. Here, we use the equalities $Q_{HV}^{\psi^+}$=$Q_{VH}^{\psi^+}$=$Q_{HV}^{\psi^-}$=$Q_{VH}^{\psi^-}$ and $Q_{HH}^{\psi^+}$=$Q_{VV}^{\psi^+}$=$Q_{HH}^{\psi^-}$=$Q_{VV}^{\psi^-}$ due to the symmetry. Here, $Q_{HH}^{\psi^+}$ can be written as
\begin{equation} \label{QHH}
\begin{aligned}
Q_{HH}^{\psi^+} =&\frac{1}{4}\sum_{i,j=0}^{\infty}P_{\mu}(i)P_{\nu}(j)Y_{ij}^{HH\psi^+},\\
Y_{ij}^{HH\psi^+}=&\sum_{p=0}^{i+j}\sum_{k=0}^{j}\Big[D_{1H}^{HH}D_{1V}^{HH}(1-D_{2H}^{HH})(1-D_{2V}^{HH})\\
&+D_{2H}^{HH}D_{2V}^{HH}(1-D_{1H}^{HH})(1-D_{1V}^{HH})\Big]P_{ij}^{HH},
\end{aligned}
\end{equation}
where $Y_{ij}^{HH\psi^+}$ is the yield, $P_{ij}^{HH}$ is the probability of quantum state $\ket{p}_{1H}\ket{0}_{1V}\ket{i+j-p}_{2H}\ket{0}_{2V}$. $D_{1H}^{HH}$ is the detection probability of detector $D_{1H}$, $1/4$ represents the probability of state $\ket{HH}$ given that the pulses sent by Alice and Bob are in the $Z$ basis, and
\begin{equation} \label{PnmHV}
\begin{aligned}
P_{ij}^{HH}&=\bigg|\frac{(-1)^{j-k}C_{i}^{p-k}C_{j}^{k}}{\sqrt{2^{i+j}i!j!}}\sqrt{p!}\sqrt{(i+j-p)!}\bigg|^2,\\
D_{1H}^{HH}&=1-(1-p_{d})(1-\eta)^p, ~D_{1V}^{HH}=p_{d}£¬\\
D_{2H}^{HH}&=1-(1-p_{d})(1-\eta)^{i+j-p}, ~D_{2V}^{HH}=p_{d},
\end{aligned}
\end{equation}
where $p_{d}$ is the dark count, $\eta=\eta_{d}\times10^{-{\beta}L/20}$ is the overall efficiency which contains channel transmittance efficiency $10^{-{\beta}L/20}$ and detector efficiency $\eta_{d}$. The distance between Alice (Bob) and Charlie is $L/2$.
Similarly, $Q_{\mu\nu}^{X}$ can be derived as
\begin{equation} \label{QCX}
\begin{aligned}
Q_{\mu\nu}^{X} &= Q_{C}^{X}+Q_{E}^{X},~~Q_{C}^{X}= 4Q_{+,+}^{\psi^+},~~Q_{E}^{X}=4Q_{+,-}^{\psi^+},
\end{aligned}
\end{equation}
where $Q_{C}^X$ and $Q_{E}^X$ represent the detection probability in the correct and false scenario, respectively.
The total quantum bit error rate of the $Z$ and $X$ basis can be written as
\begin{equation} \label{EZX}
\begin{aligned}
E_{\mu\nu}^{Z}Q_{\mu\nu}^{Z} = e_{d}Q_{C}^{Z}+(1-e_{d})Q_{E}^{Z},\\
E_{\mu\nu}^{X}Q_{\mu\nu}^{X} = e_{d}Q_{C}^{X}+(1-e_{d})Q_{E}^{X},
\end{aligned}
\end{equation}
where $e_{d}$ represents the misalignment-error probability.

We present an analytical method to acquire two tight formulas to estimate $Y_{11}^{\omega}$ and $e_{11}^{\omega}$ with only one decoy state for CSS and  $\textbf{Y}_{11}^{\omega}$ and $\textbf{e}_{11}^{\omega}$ with two decoy states (decoy+vacuum state) for non-ideal CSS. For CSS, the probability of even photon is $P_{\mu}(2i)=0$. $\mu_{1}=\nu_{1}$ is the signal state, $\mu_{2}=\nu_{2}$ is the decoy state, and $\mu_{1}>\mu_{2}>0$.
We will have
\begin{equation} \label{y1}
\begin{aligned}
\mu_{1}^4&\sinh^2\mu_{2}Q_{\mu_{2}\mu_{2}}^\omega-\mu_{2}^4\sinh^2\mu_{1}Q_{\mu_{1}\mu_{1}}^\omega=(\mu_{1}^4\mu_{2}^2-\mu_{2}^4\mu_{1}^2)Y_{11}^\omega\\
+&\sum_{j=2}^{\infty}\frac{\mu_{1}^4\mu_{2}^{m+1}-\mu_{2}^4\mu_{1}^{m+1}}{m!}Y_{1m}^\omega+\sum_{i=2}^{\infty}\frac{\mu_{1}^4\mu_{2}^{n+1}-\mu_{2}^4\mu_{1}^{n+1}}{n!}Y_{n1}^\omega\\
+&\sum_{i,j=2}^{\infty}\frac{\mu_{1}^4\mu_{2}^{n+m}-\mu_{2}^4\mu_{1}^{n+m}}{n!m!}Y_{nm}^\omega \leq (\mu_{1}^4\mu_{2}^2-\mu_{2}^4\mu_{1}^2)Y_{11}^\omega,\\
\end{aligned}
\end{equation}
\begin{equation} \label{e1}
\begin{aligned}
\sinh^2\mu_{2}E_{\mu_{2}\mu_{2}}^{\omega}Q_{\mu_{2}\mu_{2}}^{\omega}=\sum_{i,j=0}^{\infty}\frac{\mu_{2}^{n+m}}{n!m!}e_{nm}^{\omega}Y_{nm}^{\omega}\geq \mu_{2}^2e_{11}^{\omega}Y_{11}^{\omega},
\end{aligned}
\end{equation}
where we use $n=2i+1, m=2j+1$, $\{Y_{1m}, Y_{n1}, Y_{nm}\}\geq0$. So we have
\begin{equation} \label{Y11}
\begin{aligned}
Y_{11}^{\omega}& \geq \frac{(\mu_{1}^4{\sinh^2{\mu_{2}}}){Q_{\mu_{2}\mu_{2}}^{\omega}}-(\mu_{2}^4{\sinh^2{{\mu_{1}}}})Q_{\mu_{1}\mu_{1}}^{\omega}}{\mu_{1}^2\mu_{2}^2(\mu_{1}^{2}-\mu_{2}^{2})},\\
e_{11}^{\omega}& \leq \frac{\sinh^2{\mu_{2}}E_{\mu_{2}\mu_{2}}^{\omega}Q_{\mu_{2}\mu_{2}}^{\omega}}{\mu_{2}^2Y_{11}^{\omega}}.
\end{aligned}
\end{equation}
Similar to the procedure above, we have the lower (upper) bound of yield $\textbf{Y}_{11}^{\omega}$ (bit error rate $\textbf{e}_{11}^\omega$) for non-ideal CSS,
\begin{equation} \label{Y11s}
\begin{aligned}
\textbf{Y}_{11}^{\omega}\geq & \frac{P_{\mu_{1}}(1)P_{\mu_{1}}(2)g(\mu_{2})-P_{\mu_{2}}(1)P_{\mu_{2}}(2)g(\mu_{1})}{P_{\mu_{1}}(1)P_{\mu_{2}}(1)[P_{\mu_{2}}(1)P_{\mu_{1}}(2)-P_{\mu_{1}}(1)P_{\mu_{2}}(2)]},\\
g(\mu)=&Q_{\mu\mu}^{\omega}-P_{\mu}(0)Q_{\mu0}^{\omega}-P_{\mu}(0)Q_{0\mu}^{\omega}+P_{\mu}^2(0)Q_{00}^{\omega},
\end{aligned}
\end{equation}
\begin{equation} \label{Y11s}
\begin{aligned}
\textbf{e}_{11}^{\omega}\leq &\frac{1}{P_{\mu_{2}}^2(1)\textbf{Y}_{11}^{\omega}}\big[E_{\mu_{2}\mu_{2}}^{\omega}Q_{\mu_{2}\mu_{2}}^{\omega}-P_{\mu_{2}}(0)E_{\mu_{2}0}^{\omega}Q_{\mu_{2}0}^{\omega}\\
&-P_{\mu_{2}}(0)E_{0\mu_{2}}^{\omega}Q_{0\mu_{2}}^{\omega}+P_{\mu_{2}}^2(0)E_{00}^{\omega}Q_{00}^{\omega}\big].
\end{aligned}
\end{equation}

\begin{figure}[tbh]
\centering \resizebox{8cm}{!}{\includegraphics{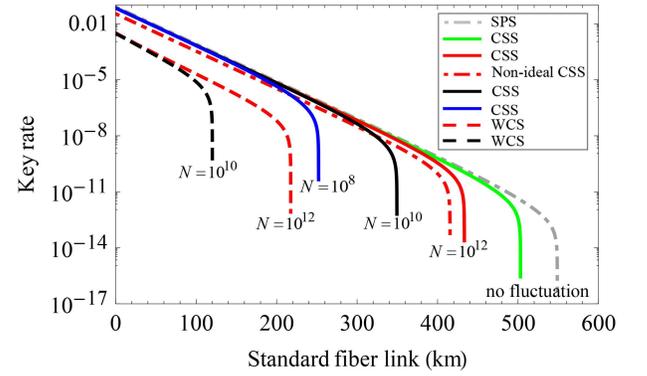}}
\caption{The secure key rates of CSS, WCS, non-ideal CSS and SPS. We use the following practical experimental parameters \cite{Tang:2014:MDIQKD}: the detection efficiency $\eta_{d}$ is 40\%,  the dark count $p_{d}$ is $1\times10^{-7}$, the intrinsic loss coefficient $\beta$ of the standard telecom fiber channel is 0.2 dB/km, the misalignment-error probability $e_{d}$ of the system is 1.5\%.}
\label{Fig:f1}
\end{figure}

To evaluate the performance of the MDI-QKD with CSS, we use the practical MDI-QKD experimental parameters \cite{Tang:2014:MDIQKD}. The single-photon state (SPS), non-ideal CSS and WCS are all used to compare with CSS. We assume that the parameter $a$ of the non-ideal CSS is 0.7, so the fidelity of the non-ideal CSS is 83.7\%. In practice, since the resources are restrained, we should consider the statistical fluctuation in finite-data case. The secure key rates shown in Fig.~\ref{Fig:f1} are estimated by the standard statistical analysis \cite{ma2012statistical}. The data length $N$ represents the number of $\mu^{\omega_{A}}\uplus\nu^{\omega_{B}}$ channel. The standard deviation is five, which implies the security bound $\epsilon=5.73\times10^{-7}$.
For WCS, in the decoy+vacuum state method, the intensities of signal state, decoy state and vacuum state are set as $\mu_{1}=0.4$, $\mu_{2}=0.07$ and 0, respectively.
While for CSS, we use one decoy state method and set $\mu_{1}=0.1$ and $\mu_{2}=0.01$. If considering non-ideal CSS, we include vacuum state.
From Fig.~\ref{Fig:f1}, we can see that the transmission distance is more than $400~km$ even with the non-ideal CSS, while the transmission distance just reaches $200~km$ with WCS in the same case.

\begin{figure}[tbh]
\centering \resizebox{8cm}{!}{\includegraphics{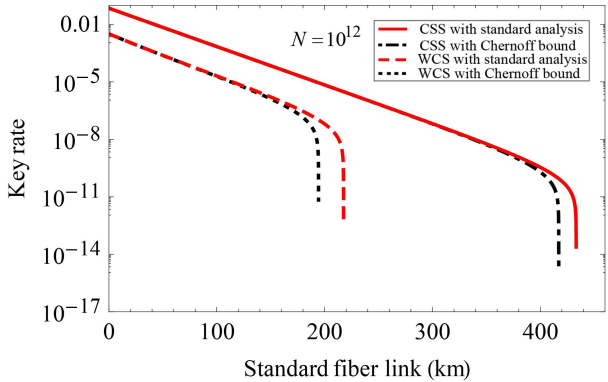}}
\caption{Comparison between standard statistical analysis and Chernoff bound. We fix the security bound to $\epsilon=2\varepsilon=5.73\times10^{-7}$. }
\label{Fig:f2}
\end{figure}

The standard statistical analysis method \cite{ma2012statistical} may not well satisfy the long distance implementation of MDI-QKD, because it is far from optimal to provide good bounds in the case of small measurement data within a reasonable time frame due to large attenuation. Therefore, we exploit a large deviation theory, Chernoff bound, to provide good bounds to estimate the parameter statistical fluctuation \cite{curty:2014:finite}. Let $X_{1},\cdot\cdot\cdot, X_{L}$ be independent Bernoulli random variables, and let $ \textbf{X}=1/L\sum X_{i}$ be the empirical mean of the variables and $E(\textbf{X})$ be the expected  value of $\textbf{X}$. So we use the following inequalities,
\begin{equation} \label{chernoff:bound}
\begin{aligned}
\Pr\left(\textbf{X}-E(\textbf{X})\geq \sqrt{2\textbf{X}\ln\left(\varepsilon^{-\frac{3}{2}}\right)}\right)\leq\varepsilon,\\
\Pr\left(E(\textbf{X})-\textbf{X}\geq \sqrt{2\textbf{X}\ln\left(16\varepsilon^{-4}\right)}\right)\leq\varepsilon,
\end{aligned}
\end{equation}
to preform the parameter estimation.
We compare the secure key rates with standard statistical analysis and Chernoff bound in Fig.~\ref{Fig:f2}.

In conclusion, we have exploited CSS as the source to improve transmission distance and secure key rate, compared with WCS in the finite-data case. We exploit the standard statistical analysis method to analyze the secure key rates of CSS, WCS and non-ideal CSS, respectively. Besides, we exploit the large deviation theory, Chernoff bound to estimate the parameter statistical fluctuation. Inspiringly, our scheme can be generalized to any photon-number distribution sources to perform MDI-QKD. With the developing technique of practical CSS, our work suggests an inspiring step towards the implementation of long distance MDI-QKD.

This work was supported by the National Fundamental Research Program under Grant No. 2011CB921300, the NNSF of China under Grant No. 61125502, the CAS and the National High Technology Research and
Development Program of China.
%%%%%%%%%%%%%%%%%%%%%%%%%%%%%%%%%%%%%%%%
% choose a style
%\bibliographystyle{ieeetr}
%\bibliographystyle{unsrt}
\bibliographystyle{apsrev4-1}
%%%%%%%%%%%%%%%%%%%%%%%%%%%%%%%%%%%%%%%%

%%%%%%%%%%%%%%%%%%%%%%%%%%%%%%%%%%%%%%%%
% choose a .bib file
%\bibliography{Bibli}
%%%%%%%%%%%%%%%%%%%%%%%%%%%%%%%%%%%%%%%%
%merlin.mbs apsrev4-1.bst 2010-07-25 4.21a (PWD, AO, DPC) hacked
%Control: key (0)
%Control: author (72) initials jnrlst
%Control: editor formatted (1) identically to author
%Control: production of article title (-1) disabled
%Control: page (0) single
%Control: year (1) truncated
%Control: production of eprint (0) enabled
%

%%%%%%%%%%%%%%%%%%%%%%%%%%%%%%%%%%%%%%%%%%%%%%%%%%%%%%%%%%%%%%%%%%%
\end{document}